\documentclass[10pt,letterpaper]{article}
\usepackage{opex3}

\begin{document}

\title{Mapping individual electromagnetic field components inside a photonic crystal}

\author{T. Denis,$^{1,*}$ B. Reijnders,$^1$ J. H. H. Lee,$^1$ P. J. M. van der Slot,$^1$ W. L. Vos,$^2$ and K.-J. Boller$^1$}

\address{$^1$ Laser Physics and Nonlinear Optics, MESA+ Institute for
Nanotechnology, University of Twente, P.O.Box 217, 7500 AE Enschede, The Netherlands\\
$^2$ Complex Photonic Systems (COPS), MESA+ Institute for
Nanotechnology, University of Twente, P.O.Box 217, 7500 AE Enschede, The Netherlands}

\email{*t.denis@utwente.nl} 


 \pagestyle{plain}
\begin{abstract}
We present a method to map the absolute electromagnetic field
strength \textit{inside} photonic crystals. We apply the method to map the electric field component $E_z$ of a two-dimensional photonic crystal slab at microwave frequencies. The slab is
placed between two mirrors to select Bloch standing waves and a subwavelength spherical scatterer is scanned inside the resulting resonator.
The resonant Bloch frequencies shift depending on the electric field at the position of the scatterer. To map the electric field component $E_z$ we measure the frequency shift in the reflection and transmission spectrum of the slab versus the scatterer position. Very good agreement is found between measurements and calculations without any adjustable parameters.
\end{abstract}




\section{Introduction}
Photonic crystals attract a tremendous deal of interest as they offer to radically control the light propagation \cite{John1987} 
and emission of light \cite{Yablonovitch1987}. In photonic crystals the dielectric constant varies periodically on
the order of the wavelength \cite{Joannopoulos2008}. Due to this periodicity, the light propagates in the form of Bloch modes \cite{Joannopoulos2008,Ashcroft1976}. An intriguing capability of photonic crystals is to
shape the local density of electromagnetic states (LDOS) inside the crystal , which is the key for
controlling the interaction of light with matter \cite{Sprik1996}. Manipulating the LDOS allows, for example,
the inhibition or the enhancement of spontaneous emission of embedded light sources [6-13]. 
This forms the basis for investigating the strong-coupling cavity regime in quantum electrodynamics
\cite{Yoshie2004}. Such manipulations also have far-reaching technological implications,
such as the development of efficient micro scale lasers, LEDs or solar cells [15-22].  

The field strength of the Bloch modes inside the photonic crystal at the locations of the emitters determine the local character of the LDOS \cite{Sprik1996}. Bloch mode fields of ideal photonic crystals, \textit{i.e.}, assuming a perfect periodicity, can be calculated by numerical methods such as finite-difference time domain (FDTD) \cite{Oskooi2010}. However, all real photonic crystals suffer inevitably from unpredictable non-periodic local deviations both due to fabrication errors and, also fundamentally, due to thermodynamical arguments \cite{Ashcroft1976,Koenderink2005}. Such deviations can not effectively be included into any kind of numerical calculations to date. A measurement is the only way to analyze the electromagnetic field inside a real photonic crystal. To eventually also characterize the LDOS of photonic crystals the field measurement method should resolve the direction and absolute value of the individual field components ($E_x$, $E_y$, $E_z$, $H_x$, $H_y$, $H_z$). A method to measure the absolute field strength of an eigenmode \textit{inside} a real photonic crystal, however, has not been reported to date.

So far the only optical method to map local fields is near-field scanning optical microscopy
(NSOM) [25-42].  
The technique relies on scanning a small tip of a tapered optical fiber \textit{above} the surface of a photonic crystal that collects part of the evanescent field with subwavelength resolution. However, NSOM suffers from
some several drawbacks. First, the method is restricted to probe local fields \textit{outside} the crystal
near its surfaces while the field deep \textit{inside} the crystal cannot be mapped. Thus one needs to make assumptions about the scattering from the internal fields under study to the detected evanescent fields. In case of three-dimensional crystals, relating the fields at the surface to the derived fields in the bulk is even more challenging \cite{Flueck2003} than in the widely studied two-dimensional slab systems.
Second, the fabrication and handling of the fiber tip is difficult to reproduce and its precise shape and
size are hard to control \cite{Burgos2003,Dickenson2007}. Hence, a pure NSOM tip probes an unknown superposition of electromagnetic field components. Third, NSOM measurements are strongly affected by a number of background effects [36-40] 
making absolute field measurements very difficult. Even in most sophisticated NSOM methods which provide nearly background free detection schemes absolute fields have not been reported \cite{Labardi2000, Esslinger2012}. What is desirable is to devise a method that can probe the absolute strength of the electromagnetic field \textit{inside} a photonic crystal. Such a method can eventually also provide a separate mapping of the individual field components, to determine the LDOS in a real photonic crystal.

Here we demonstrate a method to measure the absolute strength of the electromagnetic field distribution \textit{inside} a photonic crystal. 
Our method relies on measuring the resonant frequencies of Bloch standing waves in a photonic crystal of finite length that is enclosed by two mirrors. A subwavelength scatterer placed inside the crystal scatters the electromagnetic field which shifts the frequency of the Bloch resonances proportionally to the square of the electric and magnetic field strength at the scatterer position [43-46]. 
By measuring the frequency shifts as a function of the spatial position of the scatterer, we obtain maps of the field strengths versus position. We demonstrate the method at microwave
frequencies where fabrication errors are relatively small. Furthermore, in this frequency range the typical structure sizes are sufficiently large that a scatterer can be conveniently scanned through the crystal. To simplify the demonstration we deliberately choose a photonic crystal design where a single field component inside the microwave photonic crystal slab dominates throughout most of the crystal. Using a spherical, metallic bead as a scatterer we thus map the dominant electric field component $E_z$ \textit{inside} a photonic crystal slab.
\section{Measurement method} 
To map the electromagnetic field, a photonic crystal is placed between two mirrors. The
mirrors restrict the electromagnetic field in the photonic crystal to discrete longitudinal Bloch modes with associated resonance frequencies $\nu_0$ and wave numbers $k_z$. To measure the electromagnetic field we place a scatterer inside the photonic crystal and measure the shift of the resonance frequencies. We chose a scatterer in the Rayleigh regime with a small size parameter $x\ll1$, with $x = 2\pi R/\lambda$ . Here, $R$ is the radius of the object and $\lambda$ the
wavelength. In this regime the field is approximately constant throughout
the scatterer volume, thereby the scattering can be treated within the electrostatic approximation
to calculate the resulting frequency shift $\Delta\nu$ due to the scatterer. Using a spherical
metallic scatterer with a radius $R$ we obtain [43-45]. 
\begin{eqnarray}
	\Delta\nu(\mathbf{r})= \frac{\frac{1}{2}\mu_0\mathbf{H}(\mathbf{r})^2 - \epsilon_0\mathbf{E}(\mathbf{r})^2}{U}\pi R^3\nu_{0}
	\label{eq:shif}
\end{eqnarray}
where $\mathbf{E}(\mathbf{r})$ and $\mathbf{H}(\mathbf{r})$ are the unperturbed electric and magnetic fields at the location $\mathbf{r}$ of the
scatterer, respectively. $U$ is the total energy stored inside the unperturbed cavity, $\epsilon_0$
is the permittivity and $\mu_0$ the permeability of free space. If we measure the frequency shift versus the scatterer location $\Delta\nu(\mathbf{r})$ we will map, in a general photonic crystal, the electromagnetic field quantity $\left(\frac{1}{2}\mu_0\mathbf{H}(\mathbf{r})^2 - \epsilon_0\mathbf{E}(\mathbf{r})^2\right)$. In certain photonic crystal geometries, however, certain field components strongly dominate, hence these
components can be mapped. For instance if a crystal is designed to have the $E_z$ field dominate, \textit{i.e.}, the $E_z$ field strength is much greater than all other field components, the frequency shift becomes
\begin{eqnarray}
	\Delta\nu(\mathbf{r})\approx -\frac{\epsilon_0E_z(\mathbf{r})^2}{U}\pi R^3\nu_{0}
	\label{eq:shift_full}
\end{eqnarray}
Solving for the $E_z$ component yields 
\begin{eqnarray}
E_z(\mathbf{r})=\sqrt{\frac{-\Delta\nu(\mathbf{r}) U}{\pi R^3\epsilon_0\nu_{0}}}.
\label{eq:field}
\end{eqnarray}
Equation (\ref{eq:field}) shows that it is possible to map the absolute strength of the $E_z$ field component by measuring the
frequency shift of the longitudinal resonances $\Delta\nu$ versus the bead position and by determining the total energy stored in the cavity $U$ for a
specific input power $P_{in}$. 

\section{The photonic crystal slab} 
\begin{figure}
		\centering
		\includegraphics[width=0.98\columnwidth]{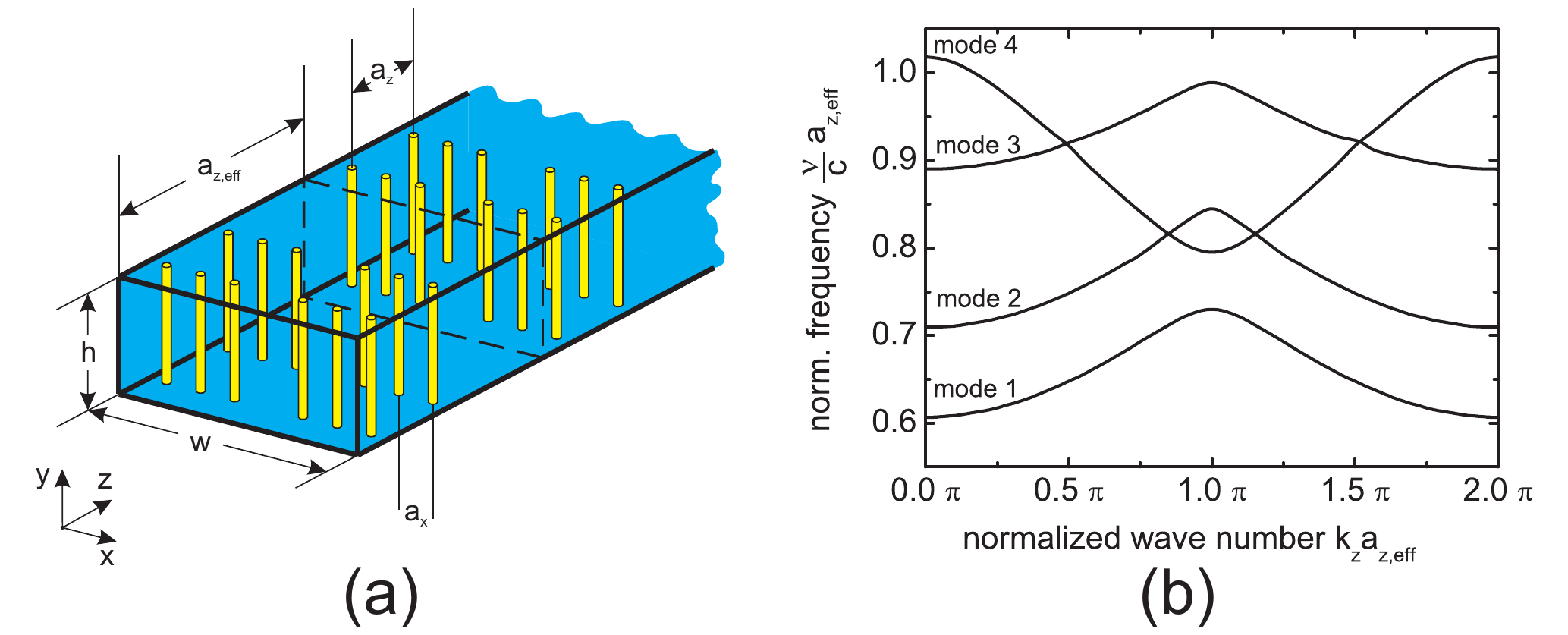}
		\caption{(a) Schematic three-dimensional view of the photonic crystal slab indicating the defining geometry parameters. Metallic rods are placed inside a rectangular metallic waveguide. The dashed line indicates the size of the supercell. (b) Calculated band structure of the photonic crystal slab showing the four lowest TE-like modes having a
non-zero $E_z$ component of the electric field.}
		\label{figure1}
\end{figure}

The unit cell of the photonic crystal slab we used (Fig. \ref{figure1}a) is designed to provide a dominant $E_z$ component in the structure, which is also required for its intended application in a photonic free-electron laser \cite{Denis2009}. The unit cell is a supercell which is surrounded by a metallic waveguide creating a two dimensional photonic crystal slab. A rectangular lattice of metal rods with a central line defect at $x=0$ forms the basis of the supercell. Along the z-direction the rectangular lattice has a lattice constant of $a_z=7.5\,\mathrm{mm}$ and the lattice constant is $a_x=6.75\,\mathrm{mm}$ along the $x$-direction. Inside the supercell the third transverse row of rods is missing. Thus the supercell consists of 12 rods and has a length of $a_{z,\mathrm{eff}\,}=22.5\,\mathrm{mm}$, indicated
in Fig. \ref{figure1}a. The diameter of the rods is 4\,mm and the surrounding waveguide has a width of
$w=47.25\,\mathrm{mm}$ and a height of $h=20\,\mathrm{mm}$. 

To calculate the band structure of the photonic crystal slab a finite-difference time-domain (FDTD)
method is used \cite{Concerto}. In the calculations the photonic crystal slab is taken to be infinitely long along
the $z$-direction by applying appropriate periodic boundary conditions to the unit cell. All metal
parts are treated as perfect electric conductors which is well justified in the microwave range. Figure \ref{figure1}b shows the results for the four lowest-frequency TE-like modes, \textit{i.e.}, modes with a non-zero longitudinal electric field component $E_z$. Due to the $z$-periodicity of the slab, the dispersion in the first Brillouin zone, \textit{i.e.},
for normalized wave numbers ($a_{z,\mathrm{eff}}\,k_z$) between 0 and $2\pi$, repeats with increasing wave number
\cite{Joannopoulos2008}. Furthermore, the finite transverse size of the waveguide results in a lowest allowed normalized frequency
of 0.61 for mode 1. No other TE-like modes exist below this cut-off frequency.
\begin{figure}
		\centering
		\includegraphics[width=0.90\columnwidth]{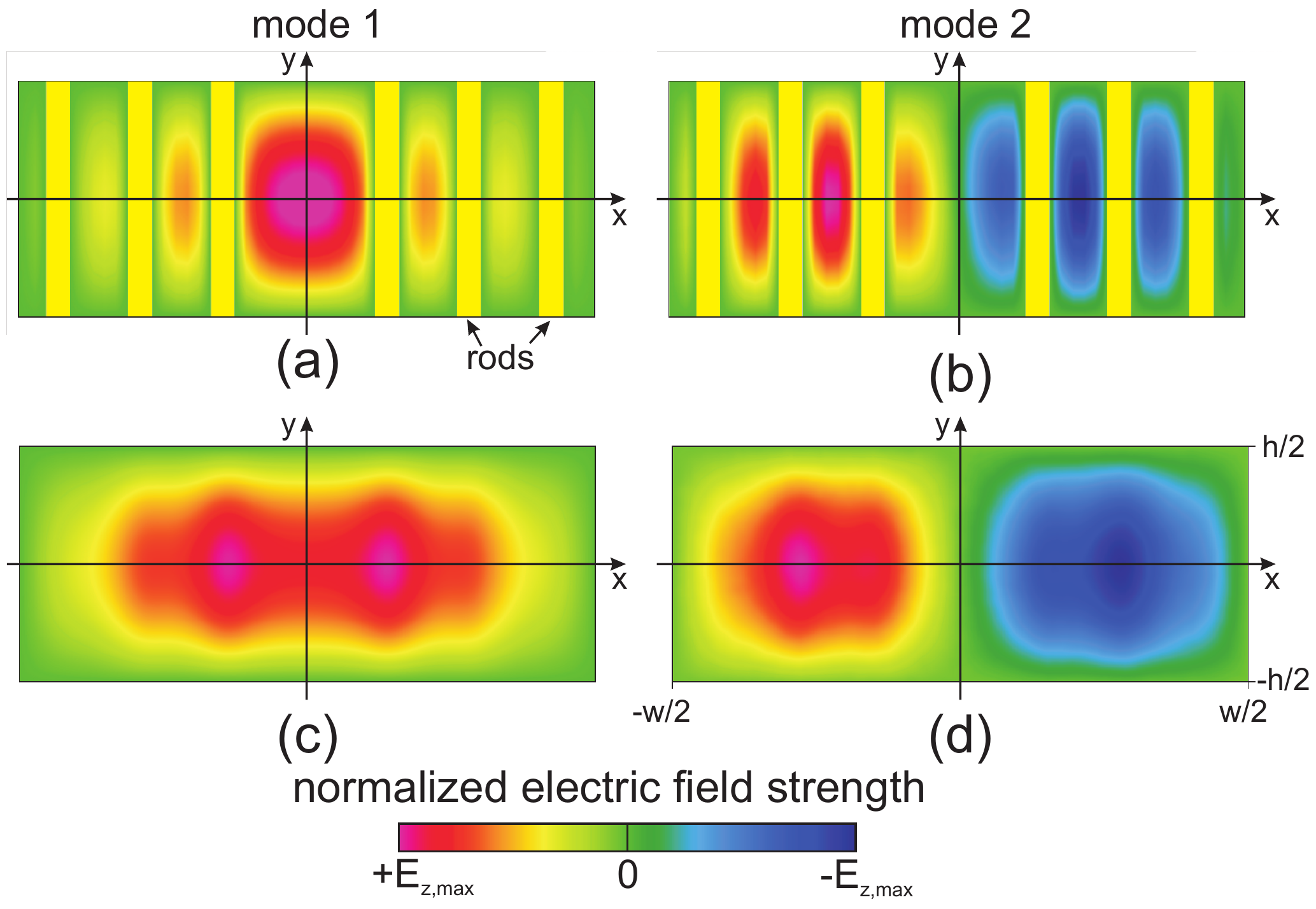}
		\caption{Transverse $E_z$-eigenmode patterns of the photonic crystal slab for mode 1 and mode 2 at two cross sections (xy-plane) inside the unit cell. First, through the first row of rods (at $z=0.5a_z$, (a) and (b)). Second, at a cross section through the empty part of the waveguide (at $z=2.5a_z$, (c) and (d)). The normalized wave number for both patterns is $k_za_{z,\mathrm{eff}}\,=0$ and the corresponding normalized frequency for (a) is 0.61 and for (b) 0.71.}
		\label{figure2}
\end{figure}

For a comparison with experimental field mapping data we calculate the local $E_z$ field distribution
of mode 1 and mode 2. The FDTD method is applied at the resonant frequency of
mode 1 and 2 for a normalized wave number of $a_{z,\mathrm{eff}}\,k_z = 0$ which corresponds to a normalized frequency of 0.61
and 0.71 respectively. Figure \ref{figure2}a through \ref{figure2}d show the $E_z$ field pattern of the modes at two
transverse planes at different $z$ coordinates. The first plane is taken through the center of the
first row of rods ($z=0.5a_z$) and the second plane is taken in the empty row of the unit cell
($z=2.5a_z$) where mapping is performed. As expected, the field pattern is symmetric due to the
symmetry of the photonic crystal slab. Furthermore, throughout all field patterns it is observed that
mode 2 has a field node at the center of the waveguide which will explain a certain effect in the experimental data later on.

\section{Experimental setup}
\begin{figure}
		\centering
		\includegraphics[width=0.75\columnwidth]{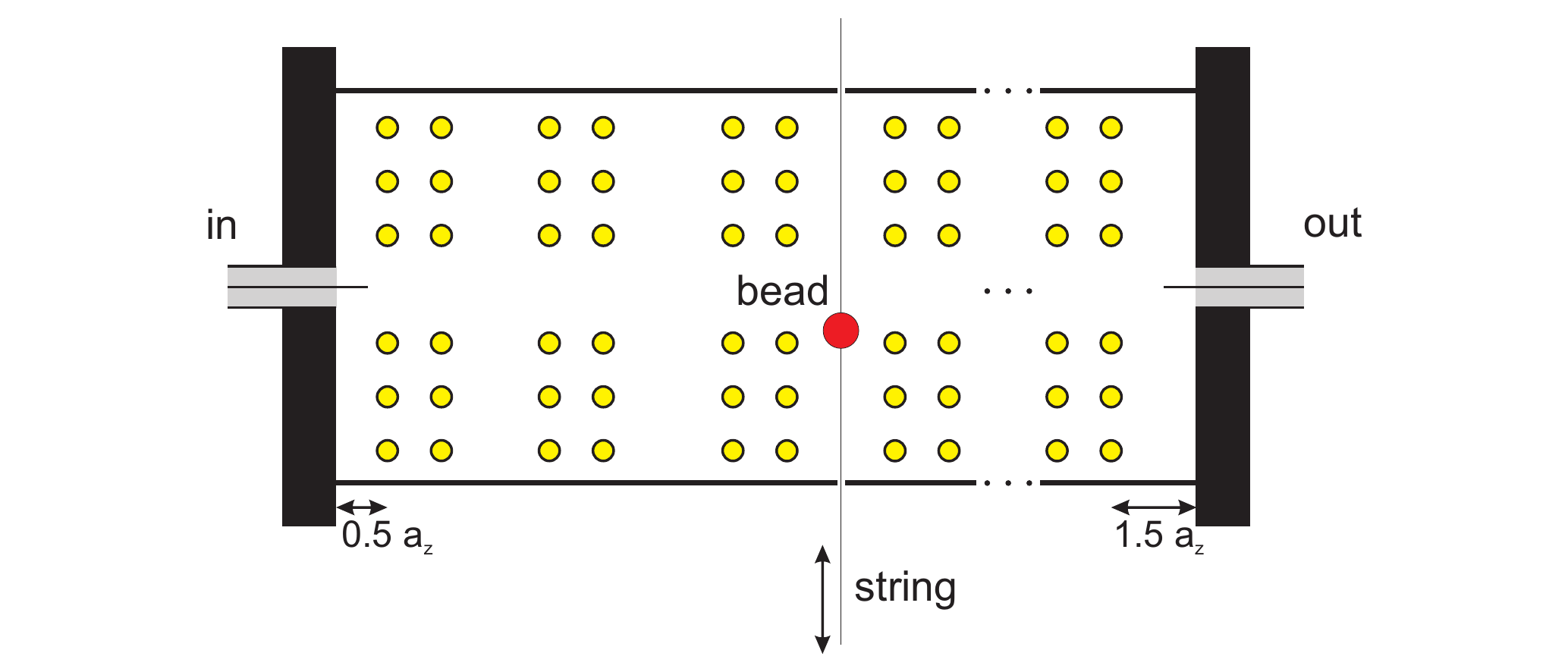}
		\caption{Schematic view of the setup to measure the electromagnetic field inside the photonic crystal slab by a scatterer. The figure shows a cross section through the photonic crystal slab	at $y=0$. The photonic crystal slab is sandwiched between two aluminum mirrors (bold black). The input and output antennas are mounted at the center of both mirrors. To map the $E_z$ field component along the $x$-direction a spherical metallic scatterer, which is mounted on a string, can be moved throughout the photonic crystal slab.}
		\label{figure3}
\end{figure}

A schematic view of the experimental setup is shown in Fig. \ref{figure3}. The photonic crystal, described in section 3, is placed between two highly reflective alumimun mirrors to form a resonator. The longitudinal resonator contains 15 unit cells of the photonic crystal in total. The first mirror is positioned at a distance of $0.5a_z$ from the center of the first row of rods and the second mirror is positioned at a distance of $1.5a_z$ from the last row of rods.

The photonic crystal slab is fabricated as follows. A channel with a rectangular cross section of
$47.25\,\mathrm{mm}\times20\,\mathrm{mm}$ is milled into a solid aluminum block (bottom plate) and is covered with
an aluminum top plate. Both parts contain holes for mounting the rods. The rods are hollow
brass cylinders with an inner diameter of 2\,mm and an outer diameter of 4\,mm. With screws
extending from the top plate through the rods into the bottom plate the rods are positioned. This
also provides an electrical connection between the rods and the waveguide. The total positioning
accuracy for the rods is estimated to a maximum of about $100\,\mu\mathrm{m}$, corresponding to a high precision of
about 1.4\% relative to the inter rod distances.

To measure the resonant frequencies of the photonic crystal slab cavity, a Hertzian dipole antenna \cite{Guru}
is mounted at the center of each mirror (see Fig. \ref{figure3}). Both antennas point along the $z-$direction
to excite or detect modes with a non-zero $E_z$ field component. One antenna acts as an emitter –
labeled ``in'' in Fig. \ref{figure3} – while the other antenna acts as a receiver for transmission measurements
– labeled ``out'' in Fig. \ref{figure3}. As a compromise between minimal loading of the resonator by the
antennae and sufficient coupling to the modes, the length of both antennas was selected as
4\,mm. Note that with the antenna position in the center of the mirror, modes with an $E_z$
field node in the center cannot be excited such as mode 2 shown in Fig. \ref{figure2}b.

The other end of both antennae is connected to a network analyzer via coaxial SMA cables.
The network analyzer is formed by a tunable microwave source with a maximum frequency
of 20\,GHz (Wiltron, model 69147A), two directional couplers (Krytar, model 2610)
and two microwave power meters (Anritsu, model ML1438A with power heads MA2444A and
MA2424B). To compensate for frequency dependent losses in cables and other components
the network analyzer is calibrated before the measurements. 
The accuracy is estimated to about 10\,\%, as re-connecting coaxial cables typically
has an effect in this magnitude. Using this setup we measure the reflection and transmission spectra of the photonic crystal slab cavity. 
For the measurements the frequency resolution is set to 250\,kHz and the input power $P_{in}$ from the network analyzer is 1\,mW.

To map the electromagnetic field inside the photonic crystal slab the scatterer is scanned through various
locations inside the resonator. The scatterer is a stainless steel bead with a radius of $R=2\,\mathrm{mm}$
which sits on a 0.3\,mm thick nylon string. As shown in Fig. \ref{figure3} the string runs through two small
holes ($300\,\mu\mathrm{m}$) in the opposing side walls. The holes are positioned in the center of the side
walls ($y=0$) and at $z=2.5a_z$ inside the 3rd unit cell, as counted from the input side. At the position of the holes in the inner surface of the side walls a 4\,mm deep
cylindrical cavity with a diameter of 4.5\,mm is fabricated in which the bead can fit completely. One end of the string is attached to a weight to keep the string straight via tension, and
the other end of the string is mounted on a translation stage. The translation stage is used
to position the bead with a relative accuracy of better than $0.05\,\mathrm{mm}$. The absolute position
is calibrated using the position at which the scatterer just completely disappears within the
cylindrical cavity, and the error in this position is estimated to be smaller than $0.1\,\mathrm{mm}$.
\section{Dispersion measurement}
Before measuring the $E_z$ field component of the photonic crystal slab resonator we verified the appropriate
description of the slab by the FDTD model. Specifically, to confirm the crystal dispersion we
measured the resonance frequencies of the different longitudinal and transverse modes without
a scatterer inside the photonic crystal. Measuring the dispersion is based on determining the longitudinal resonance frequencies of a finite-length photonic crystal consisting of $n$ unit cells and assigning wave numbers to each observed
resonance frequency \cite{Guo1992}. For the assignment we consider the phase advance $\delta\phi$ of standing waves
per round trip in the resonator. At each longitudinal resonance the phase advance per round trip
along the $z$-direction is a multiple of $\pi$:
\begin{eqnarray}
   n\delta\phi = na_{z,\mathrm{eff}}\,k_z = m\pi\qquad\qquad m={1,2...}
   \label{wave}
\end{eqnarray}
Here $L=na_z$ is the geometrical length of the resonator and $k_z$ is the wave number. As the resonator mirrors enclose 15 unit cells of the photonic crystal, $n=15$ longitudinal resonances with a finite wavelength are expected \cite{Guo1992} having a phase advance $\delta\phi$ or normalized wave number $a_{z,\mathrm{eff}}\,k_z$ of
\begin{eqnarray}
0<\delta\phi=a_{z,\mathrm{eff}}\,k_z=\frac{m}{n}\pi\leq\pi.
\label{eq:wavenumber}
\end{eqnarray}
\begin{figure}[t!]
		\centering
		\includegraphics[width=0.68\columnwidth]{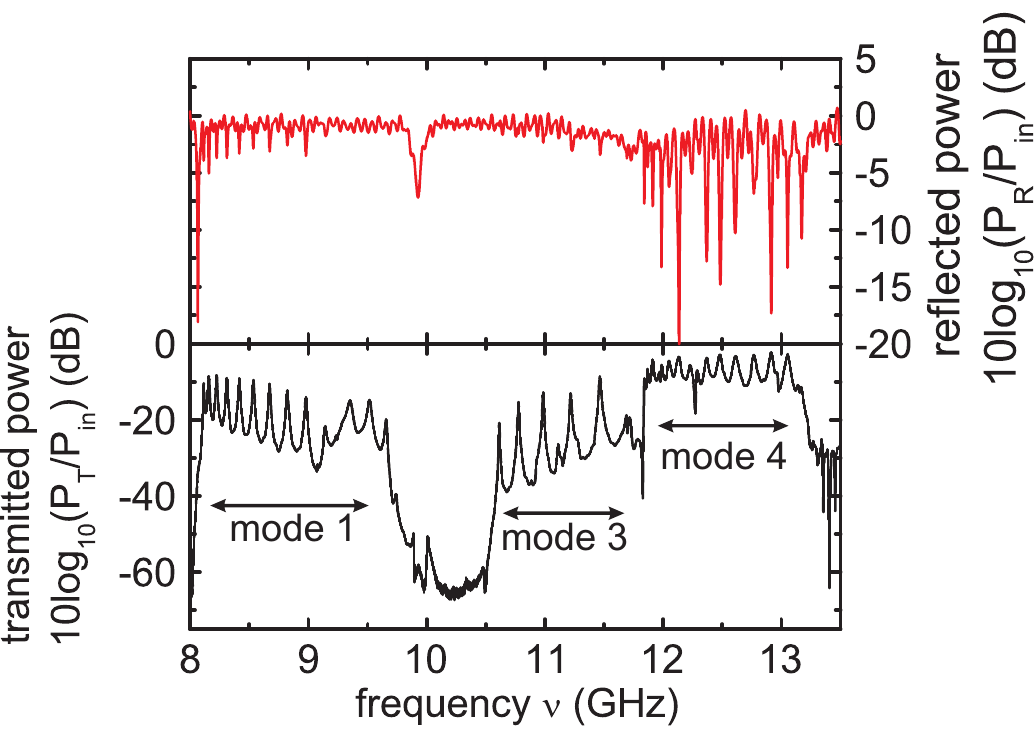}
		\caption{Transmission and reflection spectrum of the photonic crystal slab without a scatterer between 8.0\,GHz and 13.5\,GHz. The peaks correspond to the various longitudinal modes for each transverse mode.}
		\label{figure4}
\end{figure}
from the calculations shown in Fig. \ref{figure1}b the frequency of the considered modes is seen to
be a monotonously increasing or decreasing function of the wave number which renders the
resonances denumerable. By using eq. (\ref{eq:wavenumber}), a normalized wave number can be associated to
each longitudinal resonance frequency, from which the dispersion of that transverse mode is
obtained.

Figure \ref{figure4} shows the measured transmission and reflection power spectrum of the unperturbed photonic crystal slab resonator
in the range from $8\,\mathrm{GHz}$ to $14\,\mathrm{GHz}$ on a logarithmic power scale. For most of the measured frequencies the reflection is close to 0\,dB meaning that it is
equal to the input power of 1\,mW. Due to the resonator a radiation field can only build up at frequencies where longitudinal resonances of a transverse photonic crystal mode exist. In both spectra we clearly observe the resonances belonging to the longitudinal resonances.
As expected for such resonances the frequency of reflection and transmission resonances agree very well with each other.

By comparing the transmission levels of the various detected
resonances we identify which resonance belongs to the same transverse mode. Four frequency
ranges can be distinguished in Fig. \ref{figure4}. From $8.0\,\mathrm{GHz}$ to $9.8\,\mathrm{GHz}$ sharp resonances appear with an averaged
normalized transmission of about -15 dB. From $9.9\,\mathrm{GHz}$ to $10.7\,\mathrm{GHz}$ the transmission is very low
($< -40\,\mathrm{dB}$) and only one weak and broad resonance is observed. However, this is not a stop band of the slab, but a mode that cannot be effectively excited with a centered Hertzian dipole antenna due to its mode symmetry (Fig. \ref{figure2}). From $10.8\,\mathrm{GHz}$ to $11.8\,\mathrm{GHz}$ another set of resonances appears. This set of resonances is easily distinguished from the following set between $11.8\,\mathrm{GHz}$ to $13.3\,\mathrm{GHz}$ whose average transmission level ($-10\,\mathrm{dB}$) is much higher.

\begin{figure}
		\centering
		\includegraphics[width=0.98\columnwidth]{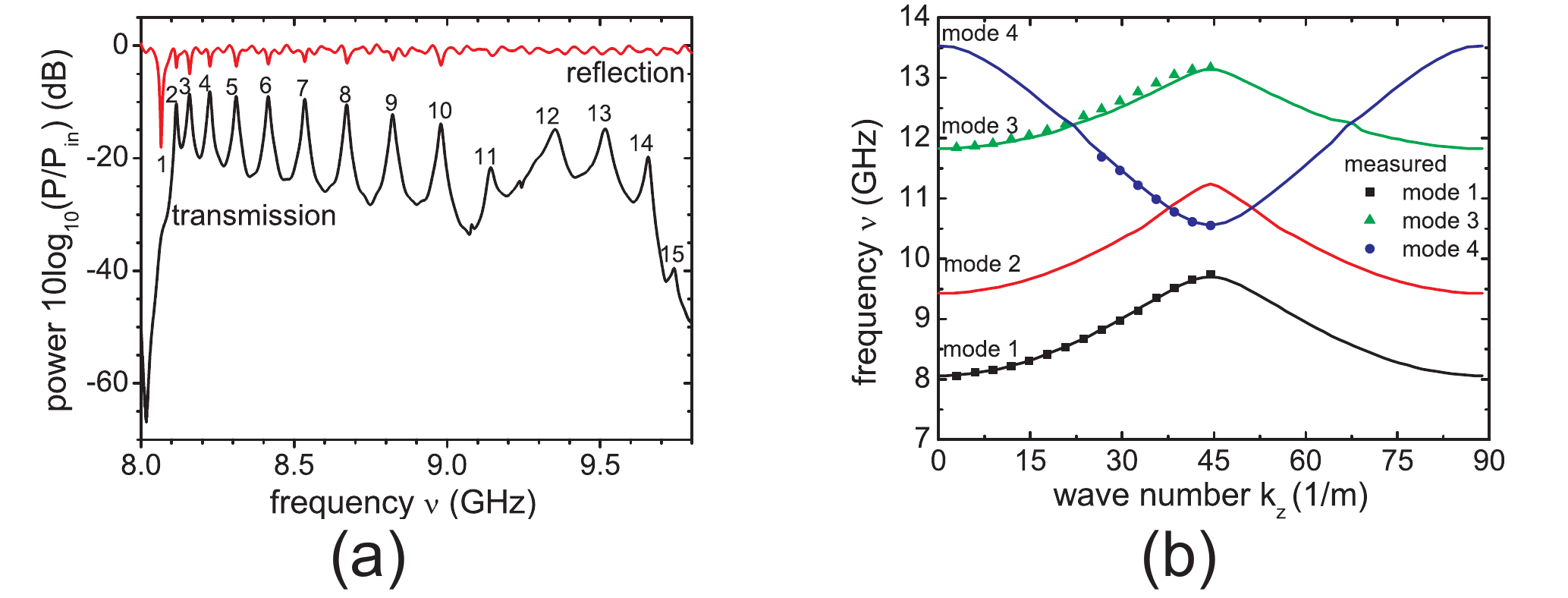}
		\caption{(a) Zoom into transmission and reflection spectrum of the photonic crystal slab depicted in Fig. \ref{figure4}. The spectra clearly show the different longitudinal modes for transverse mode 1. The labeling depicts the number of anti-nodes, $m$, along the propagation direction. (b) Measured band structure of the photonic crystal slab for the four lowest frequency TE-like eigenmodes (symbols) compared to the calculated values from Fig. \ref{figure1}b}
		\label{figure5}
\end{figure}

In total three different sets of resonances are identified as three different transverse modes.
To retrieve the dispersion of each transverse mode we assign a wave number to the longitudinal
resonances of each transverse mode. As an example for this process we concentrate on the frequency
range from $8.0\,\mathrm{GHz}$ to $9.8\,\mathrm{GHz}$. Fig. \ref{figure5}a shows a zoom into
the reflection and transmission spectrum for this frequency range. For our photonic crystal slab of 15 unit cells, we expect from theory to observe 15 longitudinal resonances
with a finite wavelength belonging to mode 1. Indeed, inspecting the transmission and reflection spectra, 15 resonances are observed. 
The first resonance is only clearly visible in the reflection spectrum. 
Furthermore, the resonances above the tenth one are
only visible in the transmission spectrum due to a lower signal-to-noise ratio in the reflected
signal.

Using these considerations we have used eq. (\ref{eq:wavenumber}) to assign wave numbers to each
measured resonance frequency of mode 1, as plotted in Fig. \ref{figure5}b. The agreement
with the theoretical dispersion (solid lines) is excellent. To assign also wave numbers
to higher order modes we take into account, as was explained above, that mode 2 is not effectively excited with
the centered Hertzian dipole antenna.  Furthermore, the resonances of mode 4 can only be partly
observed. Mode 4 overlaps in frequency with mode 3, but mode 3 couples better to the antenna
as can be seen from the higher transmission level of mode 3. Hence, only mode 3 is detected
where both modes overlap.

With these considerations we can assign normalized wave numbers to mode 3 and mode
4, as well. Also for these modes the agreement with the theoretical dispersion (solid lines) is
excellent. In conclusion, the excellent agreement for mode 1 to mode 4 indicates an appropriate
description of the fabricated slab with FDTD calculations.
\section{Electric field measurements}
We measure the longitudinal electric field inside the photonic crystal slab by measuring the frequency shift of an individual
longitudinal resonance due to a spherical metal scatterer inside the crystal. By scanning the position
of the scatterer we can map the electric field distribution. Note that the measured frequency shift $\Delta\nu$ is referenced 
to the resonance frequency of the resonator loaded by the nylon string alone. The effect of the nylon string is small as it results in a shift of only $250\,\mathrm{kHz}$. 
\begin{figure}
\centering
\includegraphics[width=0.68\columnwidth]{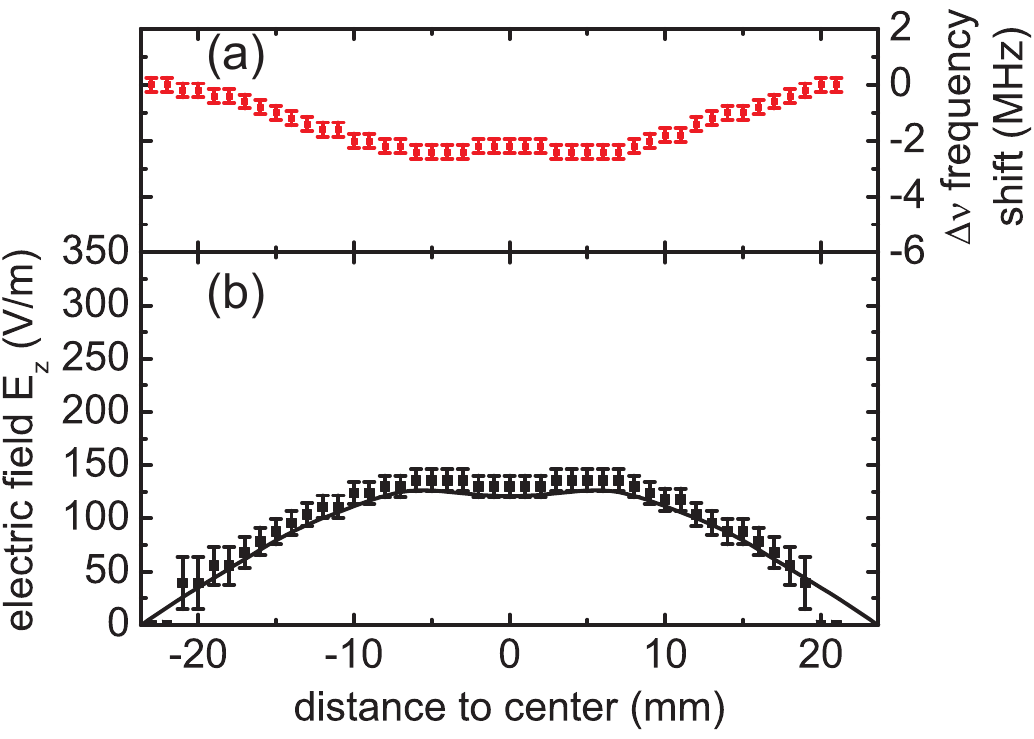}
\caption{(a) Measured frequency shift $\Delta\nu$ induced by placing the spherical scatterer at that location. Shown is the frequency shift $\Delta\nu$ for mode 1 at a longitudinal resonance with $\nu=8.16\,\mathrm{GHz}$ and $k_za_{z,\mathrm{eff}}\,=\frac{1}{15}\pi$. (b) Resulting electric field component $E_z$ from the measured frequency shift shown in (a) versus position. The measured electric field component $E_z$ (symbols) is compared to the calculated electric field (lines).}
\label{figure6}
\end{figure}

\begin{figure}
\centering
\includegraphics[width=0.68\columnwidth]{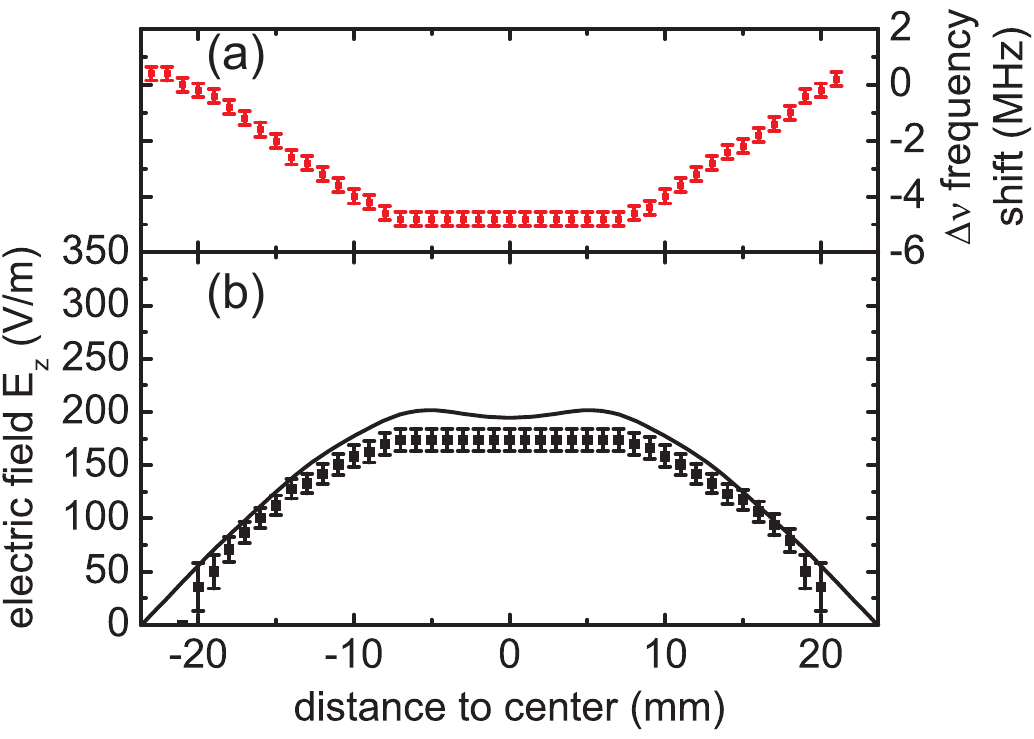}
\caption{(a) Measured frequency shift $\Delta\nu$ induced by placing the spherical scatterer at that location. Shown is the frequency shift $\Delta\nu$ for mode 1 at a longitudinal resonance with $\nu=8.42\,\mathrm{GHz}$ and $k_za_{z,\mathrm{eff}}\,=\frac{6}{15}\pi$
 (b) Resulting electric field component $E_z$ from the measured frequency shift shown in (a) versus position. The measured electric field component $E_z$ (symbols) is compared to the calculated electric field (lines).}
\label{figure7}
\end{figure}

Figure \ref{figure6}a and \ref{figure7}a show two examples of the measured frequency shift $\Delta\nu$ for mode 1 due to the scatterer. The
frequencies of the two longitudinal resonances are $8.16\,\mathrm{GHz}$ and $8.42\,\mathrm{GHz}$, respectively. The uncertainty of the frequency shift is $250\,\mathrm{kHz}$ due
to the frequency resolution of the network analyzer. In both measurements the frequency shift remains always smaller or equal to zero, as we expect from eq.(\ref{eq:shift_full}) for a mode with a dominating $E_z$ field component. Towards the edge of the waveguide the frequency shift $\Delta\nu$ approaches zero. As can be seen from Fig. \ref{figure2}, the electric field $E_z$ is at these location close to zero and consequently also the frequency shift. The strongest frequency shift of about $4.75\,\mathrm{MHz}$ and $2.25\,\mathrm{MHz}$, respectively, is reached in both examples at about $7.0\,\mathrm{mm}$ from the center. In between two rows of rods the strongest $E_z$ field component along the x-direction is generally not located at the center of the waveguide but located slightly off the center, see Fig. \ref{figure2}.

To calculate the longitudinal electric field $E_z$ from the measured frequency shift we apply eq. (\ref{eq:field}). 
However, this requires to determine the total energy stored inside the
cavity $U$. To retrieve $U$ we use the definition of the quality factor:
\begin{equation}
Q=2\pi\nu_0\frac{U}{P_{diss}}.
\end{equation}
Exactly at resonance the dissipated power per cycle $P_{diss}$ is equal to the input power $P_{in}$ from
the network analyzer. To retrieve the experimental Q-values we determine the full width half maximum of each 
transmission resonance $\nu_{FWHM}$ shown in Fig. \ref{figure5}a and use $Q=\nu_0/\nu_{FWHM}$. 

Figure \ref{figure6}b and \ref{figure7}b shows the resulting electric field strength $E_z$ (square dots) determined for the two longitudinal resonances
of mode 1 at a frequency of $8.16\,\mathrm{GHz}$ and $8.42\,\mathrm{GHz}$, respectively. For a comparison, the figure
displays also the $E_z$ field strength from the FDTD calculations presented in section 2 (black line). Uncertainties for the measured field values are determined by using the
Gaussian error propagation law. We assume that the dominating errors are the uncertainty in input power of about 10\,\% and the frequency resolution of 250\,kHz. 
All other uncertainties are much smaller and do not significantly contribute to the error bars of the measurement.

Figure \ref{figure6} shows the results for a longitudinal resonance with a frequency of $8.16\,\mathrm{GHz}$. The overall shape of the electric field shows an excellent agreement to the calculated field strength. Furthermore, also the absolute values agree within the range of the measurement accuracy, although no adjustable parameter is used in the calculations. Figure \ref{figure7} shows an example at a higher frequency of $8.42\,\mathrm{GHz}$. The agreement between FDTD calculations and experiment is again very good. This holds both for the shape of the profile and also for the absolute values. Minor deviations are visible at the waveguide center. The small difference between electric field maximum and central electric field value at the center is calculated to be about 7\,V/mm which is not resolved in the measurements. In addition, the measured absolute field value is slightly lower. We tentatively attribute both effects to the fact that at a higher frequency (shorter wavelength) influences due to non-periodic variations in the photonic crystal become more important. Nevertheless, in both examples a good agreement between experiment and theory is found. 

The two discussed examples for mapping the electric field component $E_z$ illustrate the capability of the method to map the absolute value of a
specific electric field component \textit{inside} a photonic crystal. The good agreement between the measurements and calculations further clearly demonstrates
that a field mapping \textit{inside} a photonic crystal is possible.
\section{Summary and outlook}
We have demonstrated for the first time a method for mapping the absolute strength of an
electromagnetic field component \textit{inside} a photonic crystal. The method relies on measuring the change in
resonance frequency when the photonic crystal is placed inside a resonator and the field inside the photonic crystal
is perturbed by a sub-wavelength scatterer. A spherical scatterer is applied to measure the dominating longitudinal electric field $E_z$ in a
specific photonic crystal slab. We observe a good agreement between measured and calculated electric field
strength $E_z$ without using any adjustable parameters in the calculations. Note that even if all
six components would be of comparable strength, such as in an arbitrary photonic crystal or at specific
locations, it is still possible to separately measure each field component. This can be
achieved by selecting for the scatterer a suitable material, shape and orientation such that only a
single field component contributes to the frequency shift \cite{Maier1949, Maier1952}. For example, a thin metallic
needle would short-circuit and thus probe the electric field along the orientation of the needle while leaving all
other field components unaffected.

In the future, this method could be applied in the near infrared or optical domain by scaling down the photonic crystal and the bead on a string. To induce a space dependent frequency shift a metallic or dielectric scatterer could be placed on a carbon nanotube acting as a string. To control the nanotube it could be attached to an atomic force microscope tip. Mounting a carbon nanotube on an atomic force microscope has been demonstrated \cite{Kageshima2002}. Further, atomic force microscopy allows a sufficiently high spatial resolution, demonstrated by recent measurements where an emitter directly mounted to an atomic force microscope has been used to map the emitter lifetime around a single nanorod \cite{Frimmer2011}. Combining these results it should be possible to measure the shift in resonance frequency in the transmission spectrum to determine the absolute field strength inside the air voids of a photonic crystal at optical frequencies.

\section{Acknowledgment}
This research is supported by the Dutch Technology Foundation STW, applied science division of NWO and the Technology Program of the Ministry of Economic Affairs. The authors further thank
ESA-ESTEC for providing part of the RF-equipment. The research is also part of the strategic research orientation Applied Nanophotonics within the MESA+ Research Institute.
\end{document}